\documentclass[11pt]{amsart}
\usepackage{fullpage}
\usepackage{amsmath,amsthm,amssymb,booktabs}
\usepackage{tabularx,epsfig,graphicx,url,etoolbox}

\title{Packed voters and cracked voters}

\ifdef{\SUBMIT}{}
{
\author{Gregory S. Warrington}
\address{Department of Mathematics \& Statistics, University of Vermont,16 Colchester Ave., Burlington, VT 05401, USA}
\email{gswarrin@uvm.edu}
}

\date{}

\begin{document}

\ifdef{\SUBMIT}
      {
        \baselineskip 24pt
        }{}

\newcommand{\aplan}{\mathcal{D}_0}
\newcommand{\bplan}{\mathcal{B}}
\newcommand{\cplan}{\mathcal{C}}
\newcommand{\dplan}{\mathcal{D}}

\newcommand{\bp}{\boldsymbol{p}}
\newcommand{\waste}{w}
\newcommand{\barp}{\overline{p}}
\newcommand{\noscale}[1]{\MakeLowercase{#1}}
\newcommand{\scale}[1]{\mathrm{#1}}

\newcommand{\exelec}{\mathcal{E}_0}

\newcommand{\lambdaeg}[1]{\mathrm{Gap^{#1}}}
\newcommand{\eg}{EG}
\newcommand{\dg}{DG}
\newcommand{\sg}{SG}
\newcommand{\lossg}{LG}
\newcommand{\vcg}{VCG}
\newcommand{\va}{VC1}
\newcommand{\vb}{VC2}

\newcommand{\mm}{MM}
\newcommand{\bias}{Bias}

\newcommand{\dec}{Dec}
\newcommand{\bdec}{BDec}

\newcommand{\lm}{Lop}
\newcommand{\evw}{EVW}


\newcommand{\mycapa}{Hypothetical district plan $\dplan$ for a state with 64
    voters and eight districts. Each party holds equal statewide
    support.}

\newcommand{\mycapb}{Matrix of possible transitions for a voter
  between a precursor plan $\aplan$ and a resultant district plan
  $\dplan$. Row labels indicate level of support for the voter's party
  in the voter's district in plan $\aplan$; column labels indicate
  level of support for the voter's party in the voter's district in
  plan $\dplan$. High and Very High indicate majority support; Low and
  Very Low indicate minority support. The P marks the class of packed
  voters; the C marks the class of cracked voters; and the F marks the
  class of forsaken voters.}

\newcommand{\mycapc}{Classification of voters in $\dplan$ as packed
  (\textbf{P}), cracked (\textbf{C}), forsaken (\textbf{F}), or
  neither for three possible precursor plans $\aplan$ using the
  assignments in Figure~\ref{fig:trans-table}.}

\newcommand{\mycapd}{Vote distributions for plans from
    Figure~\ref{fig:ex-pandc}. Plotted are the Dark-green Party vote
    fraction for each district, sorted in increasing order of
    support. Also shown are the value of the declination and the line
    segments used to compute its value.}

\newcommand{\mycape}{Illustration of which Republican voters in
  Maryland would be considered packed or cracked relative to two
  different comparator plans.}

\newcommand{\mycapf}{Illustration of how frequently Democratic voters
  in a given VTD as part of the 2012 congressional district plan would
  be considered to be packed/cracked relative to 1,000 randomly
  selected comparator plans. Orange (stippled) values indicate
  cracked regions with darker colors indicating higher frequencies;
  purple (solid) values indicate packed regions.}

\newcommand{\mycapg}{Illustration of how frequently Republican voters
  in a given VTD as part of the 2012 congressional district plan would
  be considered to be packed/cracked relative to 1,000 randomly
  selected comparator plans. Orange (stippled) values indicate
  cracked regions with darker colors indicating higher frequencies;
  purple (solid) values indicate packed regions.}

\newcommand{\mycaph}{Illustration of how frequently Democratic voters
  in a given ward as part of the Act 43 district plan would be
  considered to be packed/cracked relative to 1,000 randomly selected
  comparator plans. Orange (stippled) values indicate cracked regions
  with darker colors indicating higher frequencies; purple (solid)
  values indicate packed regions.}

\newcommand{\mycapi}{Illustration of how frequently Republican voters
  in a given ward as part of the Act 43 district plan would be
  considered to be packed/cracked relative to 1,000 randomly selected
  comparator plans. Orange (stippled) values indicate cracked regions
  with darker colors indicating higher frequencies; purple (solid)
  values indicate packed regions.}


\begin{abstract}
  The actions of packing and cracking are central to the construction
  of gerrymandered district plans. The US Supreme Court opinion in
  \emph{Gill \emph{v.}\ Whitford} makes clear that vote dilution
  arguments require showing that individual voters have been packed or
  cracked. In this article we provide precise definitions of what it
  means for a voter to be packed or cracked. These definitions, which
  depend crucially on the existence of at least one comparator plan,
  are illustrated using a simple hypothetical example. We also explore
  who might be considered packed or cracked for congressional plans in
  Maryland and North Carolina, and for the current state assembly plan
  in Wisconsin.
\end{abstract}

\maketitle

\section{Introduction}

A gerrymander is a district plan in which the lines have been
illicitly drawn so as to (dis)advantage one or more groups. This
rather nebulous definition is indicative of the fact that there is no
simple, concrete way to identify what is a gerrymander and what is
not. This fact is an unavoidable problem for anyone --- state courts,
federal courts, redistricting commissions, legislatures, individuals
--- attempting to scrutinize district plans for fairness. For racial
gerrymanders, the Voting Rights Act of 1965 has (at least until
recently) clarified the legal landscape in many ways. However, for
partisan gerrymandering claims, the constitutional issues are much
murkier. While the federal courts have been considering partisan
gerrymandering claims for several decades, little progress has been
made in clarifying how a successful constitutional argument against a
partisan gerrymander could be marshaled.

The case of \emph{Gill \emph{v.}\ Whitford} was the first major
partisan gerrymandering case to be reviewed by the US Supreme Court in
the last decade. The federal district court had ruled that the
Wisconsin state legislative district plan was an unconstitutional
partisan gerrymander~\cite{Whitfordgill}. However, the Supreme Court
determined that the case, as presented, depended on vote dilution
claims that require individual voters to show direct injury in order
for standing to be satisfied. To do this, a given plaintiff needed to
show that he or she had been packed or cracked.

In this article we explore what it means to show that a voter has been
packed or cracked. We begin in Section~\ref{sec:redistricting} by
reviewing the existing literature on legislative redistricting as it
pertains to partisan gerrymandering, particularly with respect to
packing and cracking. In Section~\ref{sec:pandc} we review the most
pertinent statements from the opinions in \emph{Gill
  \emph{v.}\ Whitford}. This is followed in Section~\ref{sec:ex} by
precise criteria for when a voter has been packed or cracked. As
Justice Kagan makes clear in her concurring opinion, determining
whether a voter has been packed or cracked in a given plan requires
having a comparator plan in mind. We present a simple concrete example
highlighting the dependence of the answer (i.e., on who has been
packed or cracked) on the comparator plan chosen. We close in
Section~\ref{sec:real} with various analyses of packed and cracked
voters for Maryland, North Carolina and Wisconsin. 

\section{Redistricting and partisan gerrymandering}
\label{sec:redistricting}

In this section we consider two aspects of the partisan gerrymandering
problem. The first is that of how one can identify gerrymanders. The
second is how these techniques can be applied in the service of
litigation.

\subsection{Identifying partisan gerrymanders}

The term \emph{gerrymander} derives from the salamander shape of a
Massachusetts state senatorial district signed into law by Governor
Elbridge Gerry in 1812. That the extended shape of this district was
critiqued implicitly supports the notion that regular, or ``compact,''
shapes are the most desirable and natural for districts. Indeed,
``[c]ongressionally imposed standards providing that districts be
compact, contiguous, and essentially equal in population existed
throughout most of the 19th and early 20th centuries, until
1929.''~\cite{crs}. For example, the Apportionment Act of 1911 directs
that a district be a ``contiguous and compact territory, and
containing as nearly as practicable an equal number of
inhabitants''~\cite{Apport}. While the third criterion has survived
via \emph{Wesberry \emph{v.}\ Sanders}~\cite{Wesberry} and related
cases, the first two have survived primarily at the state
level~\cite{NCSL}. (An important exception being the appearance of
compactness in the Gingles conditions arising in \emph{Thornburg
  \emph{v.}\ Gingles}~\cite{Gingles}, as applied to Section 2 of the
Voting Rights Act of 1965.)

The three properties of compactness, contiguity and equal population
are often referred to as ``traditional districting principles.''
See~\cite{Altman} for an in-depth history of the phrase; additional
desiderata such as the preservation of county boundaries or of
communities of interest are often included. As a group, these
principles provide desired and expected characteristics for
districts. It is generally accepted that any failure to follow these
principles should be justified. One approach to identifying
gerrymanders is therefore to look for violations of these traditional
principles.

The most common principle considered through this lens is that of
compactness. Distorted shapes have frequently been seen by the federal
courts as necessary characteristics of partisan
gerrymanders\footnote{``Without evidence of any distortion of
  otherwise legitimate district boundaries, there is no gerrymander,
  at least as the term is traditionally
  understood.''~\cite{Griesbach}}. However, this point of view has not
been held unanimously. Justice Souter, for example, in his
dissent~\cite{ViethSouter} in \emph{Vieth \emph{v.}\ Jubelirer} allows
for the possibility that some traditional principles might be adhered
to while others are not. Regardless, there is a large literature on
ways to measure compactness. Two of the more famous ones include the
Polsby-Popper score~\cite{PP} and the Reock score~\cite{Reock},
although dozens of other measures have been proposed. Each measure has
its pros and cons (see~\cite{compactness} for a review). Researchers
have addressed this state of affairs by either using ensembles of
measures (see, e.g.,~\cite{Azavea2010,Azavea2012} or by simply
considering multiple measures in conjunction with each other (see,
e.g.,~\cite{Hofeller,FanLi}). There are numerous examples from the
popular media in which the ``worst'' gerrymandered districts have been
identified by seeing which districts do the least well on various
(combinations of) compactness metrics (e.g.,~\cite{wp}).

There are several issues with using lack-of-compactness as the mark of
a gerrymander. The first issue is the aforementioned one that there is
a zoo of metrics one might use. A second issue is the high rate of
positives. According to one study~\cite{TwoHundred}, approximately
20\% of historical districts are less compact than the original 1812
gerrymander. Finally, contorted district boundaries should be thought
of as a \emph{symptom} of gerrymandering, rather than the mechanism by
which gerrymandering occurs.


Lack of adherence to other traditional districting principles can also
be used as evidence to support allegations of
gerrymandering. Contiguity is required for legislative districts in a
large number of states~\cite{NCSL}, though it is not required for many
congressional plans. Nonetheless, contiguity is useful only in theory
--- in practice, modern districts are all contiguous. Population
equality can, and is, used. However, the fact that the Supreme Court
has ruled in \emph{Wesberry \emph{v.}\ Sanders}~\cite{Wesberry} that
there is no \emph{de minimis} allowable deviation appears to have
consistently resulted in equipopulous maps

Scores of traditional districting principles as evidence for or
against gerrymandering are provided by the plaintiffs' brief in
\emph{Whitford \emph{v.}\ Gill}. In this brief, they compare the
enacted plan to ones from prior decades. The enacted plan \emph{does}
score worse on two compactness scores ``0.39 versus 0.41, and 0.28
versus 0.29,''~\cite[pg. 36]{Gillbrief}, but the differences are
modest. On the other hand, computer simulations are utilized in the
same brief to argue that the Act 43 plan is an outlier with respect to
the number of split municipalities~\cite[Figure 11,
  pg. 40]{Gillbrief}.

In recent decades there has been significant interest in developing
metrics specifically tailored to identifying gerrymandering. Many of
these metrics are based on finding asymmetries in the
\emph{seats-votes curve}. This curve, studied since at
least~\cite{tufte}, relates the fraction of the statewide vote each
party gets to the fraction of seats it wins. So, for example, the
curve might pass through the points (0.3,0.2) and (0.5,0.45) if
winning 30\% or 50\% of the statewide vote would lead the Democrats to
winning 20\% or 45\%, respectively, of the total seats
available. Notwithstanding the fact that a number of assumptions must
be made in order to compute the seats-votes curve, several partisan
gerrymandering metrics have been derived from it. Gelman and
King~\cite{GK} suggest partisan bias, which considers the fraction of
seats each party wins under the assumption that they win 50\% of the
statewide vote. The mean-median bias, which has been advocated for by
a number of authors in various forms~\cite{Nagle1, Wang,
  McDonald-two}, returns a related deviation. Other aspects of the
seats-votes curve such as responsiveness and overall competitiveness
have also been considered as means of identifying gerrymanders. More
recent suggestions that are less directly related to the seats-votes
curve include the efficiency gap~\cite{McGhee} (and many variations;
see~\cite{Comparison} for a summary) and the
declination~\cite{declination}.

\subsection{Litigation}

The aforementioned measures are simple mathematical functions. They
consider a limited amount of electoral and/or redistricting data and
attempt to provide insight into the extent to which a district plan is
unfair. The first real test for such measures arrived in in 1986 in
\emph{Davis \emph{v.}\ Bandemer}. Indiana Democrats argued that the
1981 Indiana legislative district plan violated the Equal Protection
Clause (EPC) of the 14th amendment. The court determined that
``political gerrymandering\ldots is properly justiciable under the
Equal Protection Clause,''~\cite{Davis}. However, the justices were
unable to agree on a standard. 

The next significant partisan gerrymandering case was \emph{Vieth
  \emph{v.}\ Jubelirer}. In this case, the 2001 Pennsylvania
congressional plan was challenged on both EPC grounds as well as on
First Amendment grounds. There was little consensus in the court's
ruling; a plurality of four justices determined such cases are
nonjusticiable, four found them justiciable and proposed various
standards, and Justice Kennedy wrote that such cases are justiciable
but was not satisfied with any yet proposed standard. In 2006, the
court heard \emph{LULAC \emph{v.}\ Perry} regarding alleged violations
of the EPC and VRA in relation to mid-decade congressional
redistricting in Texas. In the opinion of the court, written by
Kennedy, some support was shown for partisan asymmetry metrics, in
particular partisan bias, but ultimately the court determined that no
one had yet proposed a suitable, manageable standard.

Central to any manageable standard is the legal argument on which it
is based. For racial gerrymandering cases, while the details are
intricate, the general legal argument is straightforward. As described
in~\cite{tokaji}: ``The U.S. Supreme Court has issued three major
decisions on [racial gerrymandering] since 2015. All these cases\ldots
[are] challenged under the Equal Protection Clause on the ground that
they packed racial minorities into districts in a way that was not
justified by the interest in complying with the Voting Rights Act
(VRA).'' For partisan gerrymandering, the best approach has been less
clear. Not surprisingly, success in the courts for partisan cases has been,
correspondingly, much more limited.

In the subsequent decades, a number of proposals have been put forth
for what such a manageable standard should look like. Recent examples
include~\cite{M-S,Wang,grofmanStd}. Contemporaneously, several
partisan gerrymandering cases have been wending their ways through the
federal court system. 

The Supreme Court heard two partisan gerrymandering cases in
2017--2018. While the court reviewed \emph{Benisek
  \emph{v.}\ Lamone}~\cite{Benisek}, it did not address justiciability of
partisan gerrymandering claims. However, a more substantive response
resulted from \emph{Whitford \emph{v.}\ Gill}~\cite{Gillwhitford}. In
this case, the Wisconsin state legislative plan was challenged on both
EPC and First Amendment grounds. In its opinion striking down the
plan, the federal district court focuses its discussion of standing on
the EPC. The Supreme Court, in its review of the case, adjudged that
the plaintiffs did not, in how it presented the case, show the
``injury in fact'' necessary for Article III standing. Both Chief
Justice Roberts, writing for the court, and Justice Kagan, in a
concurring opinion, described what was necessary to show the injury
required. We turn to a discussion of those remarks in the next
section.

\section{Packing and cracking in \emph{Gill \emph{v.}\ Whitford}}
\label{sec:pandc}

The problem of recognizing a partisan gerrymander reduces to that of
recognizing the presence of packing and cracking: ``For packing and
cracking are the ways in which a partisan gerrymander dilutes votes''
(\emph{Gill \emph{v.}\ Whitford,} concurring opinion). This fact
necessitates a careful definition of these terms.

A widely cited definition for packing and cracking is the following
given by Justice Scalia~\cite{ViethScalia}:
\begin{quotation}
  ```Packing' refers to the practice of filling a district with a
  supermajority of a given group or party. `Cracking' involves the
  splitting of a group or party among several districts to deny that
  group or party a majority in any of those districts.''
\end{quotation}
Unfortunately, this definition is not consistent in its treatment of
the two actions. As defined by Justice Scalia, packing happens
whenever a supermajority arises, regardless of the circumstances;
cracking only arises in the context of intent. There are two obvious
modifications one could apply to make these definitions consistent
with each other.

One option is to remove the intent criterion from cracking. However,
this leaves us with definitions for packing and cracking that are less
directly connected to the creation of partisan gerrymanders. (Some
locales are primarily populated by the supporters of a single party. A
supermajority could easily arise without any attempt to create a
gerrymander.) As a consequence, we prefer to modify the definition of
packing to be the following: ``Packing'' is the practice of filling a
district with extra members of a given group or party so as to prevent
those extra members from contributing to majorities in other
districts.

Under these modified definitions, the recognition that packing or
cracking has occurred contains an implicit acknowledgment that
different choices could have been made. Consequently, to show that
cracking or packing has occurred, one should be able to point to an
acceptable, alternative district plan in which a reconfiguration of
districts leads to additional majorities by the disadvantaged group or
party. This is essentially the procedure Justice Kagan describes in
\emph{Whitford \emph{v.}\ Gill} (concurring opinion):

\begin{quotation}
``For example, a Democratic plaintiff living in a 75\%-Democratic
district could prove she was packed by presenting a different map,
drawn without a focus on partisan advantage, that would place her in a
60\%-Democratic district. Or conversely, a Democratic plaintiff
residing in a 35\%-Democratic district could prove she was cracked by
offering an alternative, neutrally drawn map putting her in a 50-50
district. The precise numbers are of no import. The point is that the
plaintiff can show, through drawing alternative district lines, that
partisan-based packing or cracking diluted her vote.''
\end{quotation}

We note that the above argument implies that voters suffer harm merely
by being placed in less competitive districts.

In our above discussion, we have only considered packing and cracking
as actions. From such a viewpoint, one is naturally provided with at
least two district plans: a starting district plan $\aplan$ and a
district plan $\dplan$ that is the result of the packing and/or
cracking. In this situation, which districts have been packed or
cracked follows directly from where the packing and cracking have
taken place. However, the designation is most cleanly described in
terms of districts of the original plan $\aplan$. As we discuss in
Section~\ref{sec:ex}, there are some ambiguities regarding how to
label the districts in $\dplan$. As such, the references to packed and
cracked districts in the following quotation from Kagan's \emph{Gill
  \emph{v.}\ Whitford} concurring opinion must be considered
carefully.

\begin{quotation}
``Consider the perfect form of each variety. When a voter resides in a
packed district, her preferred candidate will win no matter what; when
a voter lives in a cracked district, her chosen candidate stands no
chance of prevailing. But either way, such a citizen's vote carries
less weight --- has less consequence --- than it would under a
neutrally drawn map.''
\end{quotation}

Another problem with the phrase ``packed district'' is that it
obscures the dependency on a precursor district plan. So, for example,
Chief Justice Roberts' majority opinion in \emph{Gill \emph{v.}\ Whitford}
makes the situation sounds simpler than it actually is:

\begin{quotation}
``To the extent the plaintiffs' alleged harm is the dilution of their
votes, that injury is district specific. An individual voter in
Wisconsin is placed in a single district. He votes for a single
representative. The boundaries of the district, and the composition of
its voters, determine whether and to what extent a particular voter is
packed or cracked.''
\end{quotation}

The dilution does not depend \emph{only} on the district and the
voters in it. It also depends on the suite of comparator plans being
used to determine which districts have been packed or cracked.

In the next section we shall address what it means for a voter to be
packed or cracked. (The case of districts will be treated briefly.)
As illustrated by the above \emph{Gill \emph{v.}\ Whitford} excerpts,
identifying voters as having been packed or cracked is what is needed
for vote dilution claims.

\section{Packed and cracked voters}
\label{sec:ex} 

In this section we explore what it means for individual voters to be
packed or cracked. We will do so in the context of a hypothetical
state with 64 voters living in an $8\times 8$ grid as shown in
Figure~\ref{fig:ex-single}. The Dark-green Party and the Light-gray Party each
enjoy equal statewide support. We have illustrated in the figure an
undeniably compact district plan $\dplan$ consisting of eight
districts. From the vantage point of seats-votes proportionality, this
appears to be an unfair district plan. Even though statewide support
for the two parties is equal, the Dark-green Party wins only two of the
eight seats. And indeed, partisan asymmetry metrics such as the
efficiency gap~\cite{McGhee} and the declination~\cite{declination}
agree with this intuition (see Figure~\ref{fig:prec-vd}.D). But it is
not clear \emph{which} districts have been packed or cracked. The
analogous questions for individual voters are not any clearer. As
discussed in Section~\ref{sec:pandc}, such determinations require a
comparator plan.

\ifdef{\SUBMIT}{}
{
\begin{figure}
  \centering
  \includegraphics[width=0.25\linewidth]{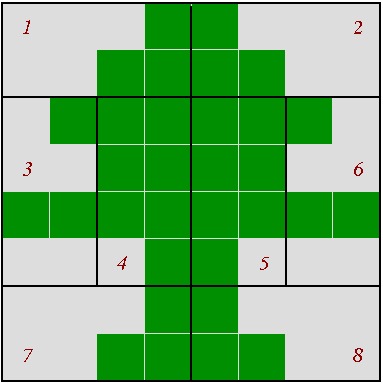}
  \caption{\mycapa}
  \label{fig:ex-single}
\end{figure}
}

We emphasize the necessity of a comparator plan by considering some of
the districts and voters in Figure~\ref{fig:ex-single} in
detail. Consider, for example, the Dark-green voters in the overwhelmingly
Dark-green Districts 4 and 5. The Dark-green Party's overwhelming support in
these two districts certainly isn't very efficient, but if these
districts cover a community of interest, then it may be appropriate. A
verdict that these districts were packed requires the demonstration
of, at the least, an acceptable plan in which similar districts have
lower proportions of Dark-green voters.  Similarly, consider the
minority Dark-green District 2. Modest changes to its boundary could make
it a majority Dark-green district. But describing one of the Dark-green voters
in this or a nearby district in $\dplan$ as having been cracked
presupposes that there is a reasonable district plan --- and hence a
reasonable district --- with those voters in the majority in that
single district. Without the explicit demonstration of such a plan,
one must allow for the possibility that external constraints or
priorities make District 2 defensible as part of a fair district
plan. An analogous argument applies when considering whether an
individual voter was cracked.

Now assume that we have identified a fair precursor plan, $\aplan$. We
can suggest at least tentative definitions for what it means for a
voter to be packed or cracked in $\dplan$ relative to $\aplan$ as
follows. To do so, without loss of generality we focus on Dark-green Party
voters. First note that four classes of Dark-green voters naturally arise
by considering whether a given voter starts in a majority or
minority Dark-green district in $\aplan$ and whether they end up in a
majority or minority Dark-green district in $\dplan$. For voters who
remain in a majority Dark-green district or remain in a minority Dark-green
district, we obtain a finer classification by considering whether the
support for the Dark-green Party in the given voter's district goes up or
down in the transition from $\aplan$ to $\dplan$. The resulting
transitions are summarized in Figure~\ref{fig:trans-table}.

\ifdef{\SUBMIT}{}
{
\begin{figure}
  \centering
  \includegraphics[width=0.3\linewidth]{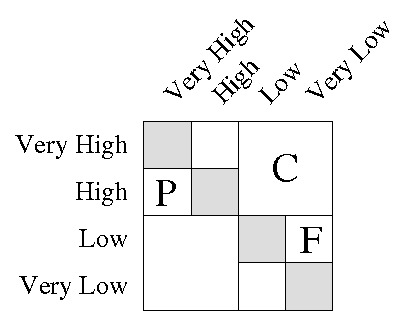}
  \caption{\mycapb}
  \label{fig:trans-table}
\end{figure}
}

With the universe of possible transitions now identified, we are now ready
to explore which Dark-green voters should be identified as having been
packed or cracked. We first consider those transitions that do not
correspond to any form of packing or cracking. Four of the transitions
are completely uninteresting: A voter starts and ends in a district
with the same level of Dark-green Party support. These are indicated in
Figure~\ref{fig:trans-table} by the grayed-out squares. Next we note
that there are a number of redundancies when a voter transitions from
a majority Dark-green district to a minority Dark-green district or vice
versa. For example, it does not make sense for our purposes to
distinguish a voter transitioning from a majority Dark-green district
to a Low Dark-green district from a voter transitioning from a
majority Dark-green district to a Very Low Dark-green
district. These redundancies are indicated in
Figure~\ref{fig:trans-table} by the $2 \times 2$ squares. For the lower
left square, the voter transitions from a district in which the Dark-green
Party candidate loses to one in which she wins; it is hard to see how
this voter could be construed as having been packed or cracked.

The final two blank squares in Figure~\ref{fig:trans-table} correspond
to transitioning from a Very High Dark-green district to a High
Dark-green district or from Very Low Dark-green district to Low
Dark-green district. It is hard to interpret either as a harm to the
voter as an individual. In both cases, he remains in the same type of
district in which he started --- either the Dark-green Party candidate
wins or she doesn't. And in each case, the voter's vote is, if
anything, \emph{more} likely to be consequential. In the Very
High-to-High case, the Dark-green support in the district is lessened
and the given voter's support can be viewed as even more important
than before. In the Very Low-to-Low case, the Dark-green candidate
still loses, but had things been slightly different, perhaps the given
voter's vote would have been the one to push the Dark-green candidate
to victory. Irrespective of how efficiently or inefficiently the
Dark-green Party's votes are being utilized in this district, the
individual voter certainly shouldn't feel that his support is less
valuable in $\dplan$ than in $\aplan$.

We are left with three types of transitions to classify. Transitioning
from High to Very High is consistent with the notion of a voter being
``packed.'' This case is denoted by a ``P'' in
Figure~\ref{fig:trans-table}. Similarly, transitioning from majority
Dark-green to minority Dark-green is consistent with the voter being
cracked; this is denoted by a ``C'' in
Figure~\ref{fig:trans-table}. The final case occurs when a voter
transitions from a Low Dark-green district to a Very Low Dark-green
district. This is in some sense analogous to what happens to the
packed voter, except that the voter's candidate of choice suffers a
\emph{worse loss} rather than a stronger win. As such, we consider
this transition a disadvantage for the voter; we denote this class of
voters by an ``F'' for ``forsaken.'' While Dark-green voters are
likely to be packed or cracked when the Light-gray Party is creating
the gerrymander, it is Light-gray voters who are more likely to be
forsaken in such a scenario: Forsaking Dark-green voters would result
in the distribution of additional Dark-green voters into other
districts the Light-gray party was trying to win. As suggested by this
last observation, the classification in Figure~\ref{fig:trans-table}
applies, \emph{mutatis mutandis}, equally well to both Light-gray
voters and Dark-green voters. Finally, we note that packed voters of
one party will, locally, be surrounded by forsaken voters of the other
party and vice versa.

Now that we know how to classify individual voters, we are ready to
see how voters of Figure~\ref{fig:ex-single} get classified under
various precursor plans. This is illustrated in
Figure~\ref{fig:ex-pandc} for the three precursor plans $\aplan$
depicted in the first row. The second row consists of three copies of
$\dplan$, each indicating the corresponding packed/cracked/forsaken
voters according to the given precursor plan as determined by
Figure~\ref{fig:trans-table}. As should be clear from the example, the
classification of any individual voter in $\dplan$ as being packed,
cracked, forsaken or neither is highly dependent on the precursor plan chosen.

\ifdef{\SUBMIT}{}
{
\begin{figure}
  \centering
  \includegraphics[width=0.7\linewidth]{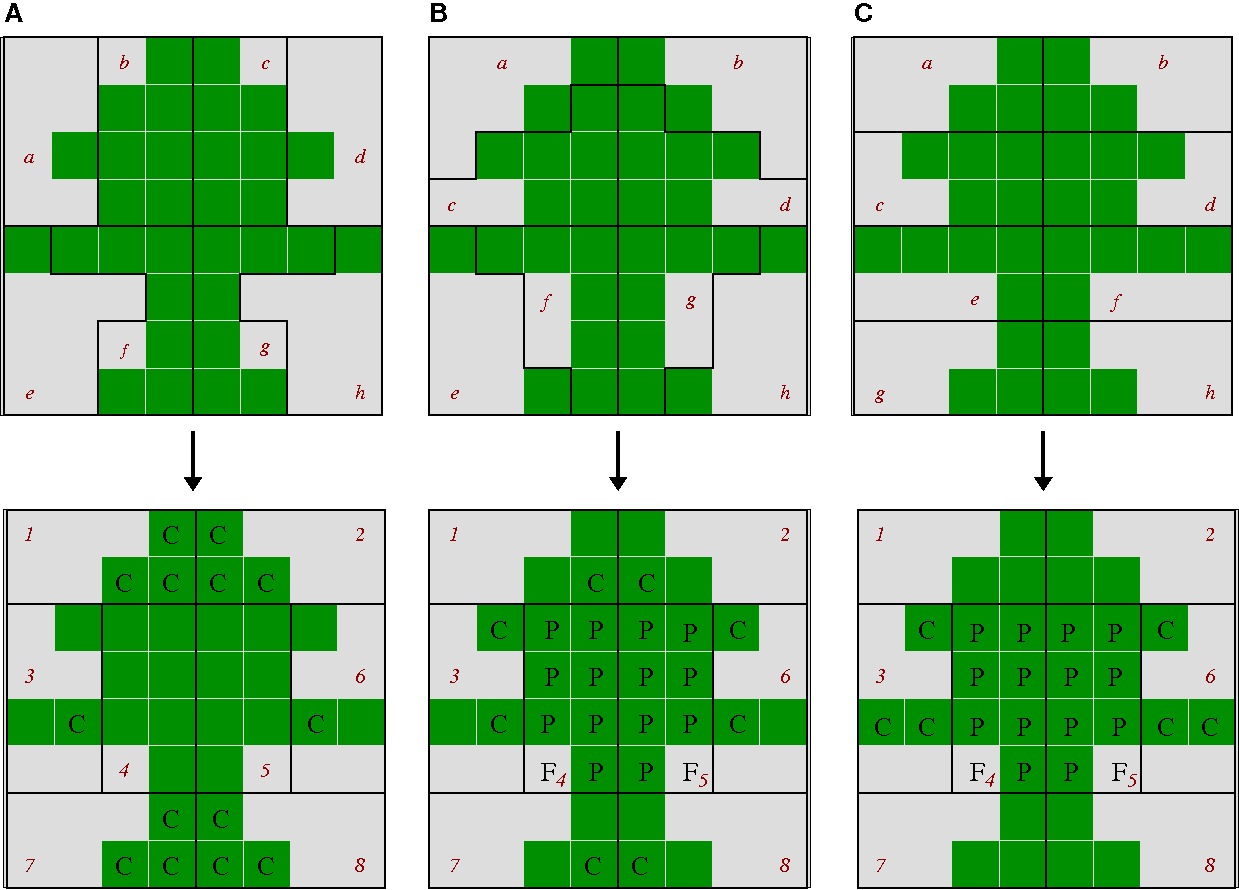}
  \caption{\mycapc}
  \label{fig:ex-pandc}
\end{figure}
}

We now return to the matter of identifying which districts in the
plan $\dplan$ have been packed or cracked. It follows from the
definitions of packed and cracked voters that a single district in
$\dplan$ cannot contain \emph{both} packed voters and cracked voters:
A Dark-green packed voter must end up in a majority Dark-green district while
a cracked voter must end up in a minority Dark-green district. So one
possibility is to define a packed (cracked) district in $\dplan$
relative to $\aplan$ as a district containing at least one packed
(cracked) voter. For example, in Figure~\ref{fig:ex-pandc}.A,
there are six cracked districts (1, 2, 3, 6, 7 and 8) in $\dplan$ and
no packed districts while in Figure~\ref{fig:ex-pandc}.C there
are two packed districts (4 and 5) and two cracked districts (3 and
6).

There is another approach to identifying packed and cracked districts
that is in some ways more natural, but depends on a (partial) matching
between districts of $\aplan$ and those of $\dplan$. It is not
necessarily consistent with the above, bottom-up approach. It works by
applying the transition matrix of Figure~\ref{fig:trans-table}
directly to districts rather than descending down to the level of
voters. So, for example, in Figure~\ref{fig:ex-pandc}.A, if Districts
\emph{c} and 2 are paired, then District 2 would qualify as a cracked
district since District $c$ is majority Dark-green while District 2 is
not. However, if District 2 is instead paired with District \emph{d},
then District 2 would not qualify as cracked.

We end this section by illustrating, in Figure~\ref{fig:prec-vd}, the
vote distributions for the various district plans depicted in
Figure~\ref{fig:ex-pandc} along with the values of one of the partisan
asymmetry metrics, the declination (introduced in~\cite{declination}). The
symmetry of the first three plans illustrates that the plans we have
used for $\aplan$ treat the parties symmetrically. In the fourth plan,
one can see that the Dark-green Party loses six close districts while
winning two districts overwhelmingly.

\ifdef{\SUBMIT}{}
{
\begin{figure}
  \centering
  \includegraphics[width=1\linewidth]{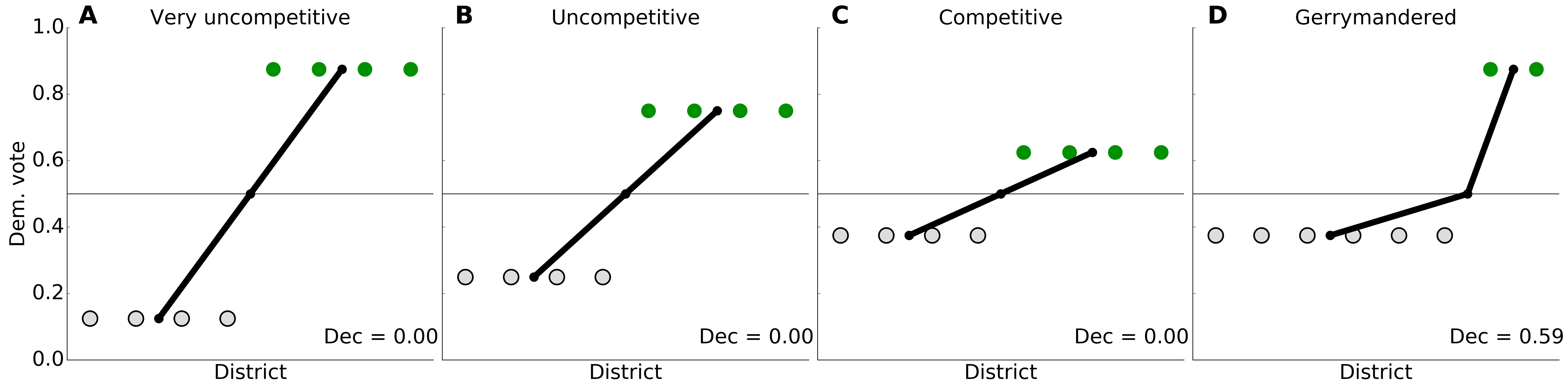}
  \caption{\mycapd}
  \label{fig:prec-vd}
\end{figure}
}

\section{Packed and cracked voters in Maryland, North Carolina and Wisconsin}
\label{sec:real}

In this section we use recent electoral data to explore which voters
have been packed or cracked in the states of Maryland, Wisconsin and
North Carolina. Before we do so, it is worth elaborating on how
forsaken voters are related to packed voters.

Consider an atomic geographic unit such as a ward as it relates to a
comparator plan $\aplan$ and a final plan $\dplan$. Suppose the
Dark-green voters in the ward are packed relative to these two
plans. This means that the ward lies in a Dark-green majority district
in both plans and that the proportion of Dark-green voters is higher
in $\dplan$. This characterization is equivalent to saying that the
ward lies in a Light-gray minority district in both plans and that
the proportion of Light-gray voters is lower in $\dplan$. As such, any
Light-gray voters in this ward are automatically forsaken. In light
of this connection, we will refrain from referring explicitly to
forsaken voters in the following analyses. Their presence for one
party can be immediately inferred from the presence of opposing-party
packed voters.

\subsection{Maryland}
\label{sec:md}

In \emph{Benisek \emph{v.}\ Lamone}, Republican voters challenged the
Maryland 2011 congressional plan on the grounds that Republican voters
of the 6th congressional district had been cracked. We investigate
this claim by pairing the district plan with two different precursor
plans. The plans we use were generated as part of the \emph{Atlas of
  Redistricting Project}~\cite{atlas}. The first comparator plan we
use was generated with the goal of making each district as competitive
as possible while still respecting county boundaries. In
Figure~\ref{fig:md}.A we illustrate this plan with each district
colored according to how strongly it leans Democratic or
Republican. The \emph{Atlas} project computed this lean using the Cook
Political Report's Partisan Voter Index (PVI) and we use their values
here. The competitiveness of the districts is borne out by the light
shading of each district. In Figure~\ref{fig:md}.E we have displayed
the current Maryland congressional plan with regions colored according
to PVI values taken from~\cite{PVI}. The high saturation levels are
indicative of relatively uncompetitive districts. In the middle, as
Figure~\ref{fig:md}.C, we illustrate the regions of the state
containing Republican voters who were packed or cracked in the enacted
plan relative to the competitive plan. In support of the plaintiffs'
claims, much of the Western part of state (i.e., the 6th district) is
shown in orange as a result of those voters having been cracked. This
is consistent with the fact that the 6th (westernmost) district is
slightly Republican leaning in the competitive plan, but is Democratic
leaning in the enacted plan. The comparison indicates packing of
Republicans in the 1st (easternmost) district as well as some cracking
centered on the 2nd district in the vicinity of Baltimore.

\ifdef{\SUBMIT}{}
{
\begin{figure}
  \centering
  \includegraphics[width=1\linewidth]{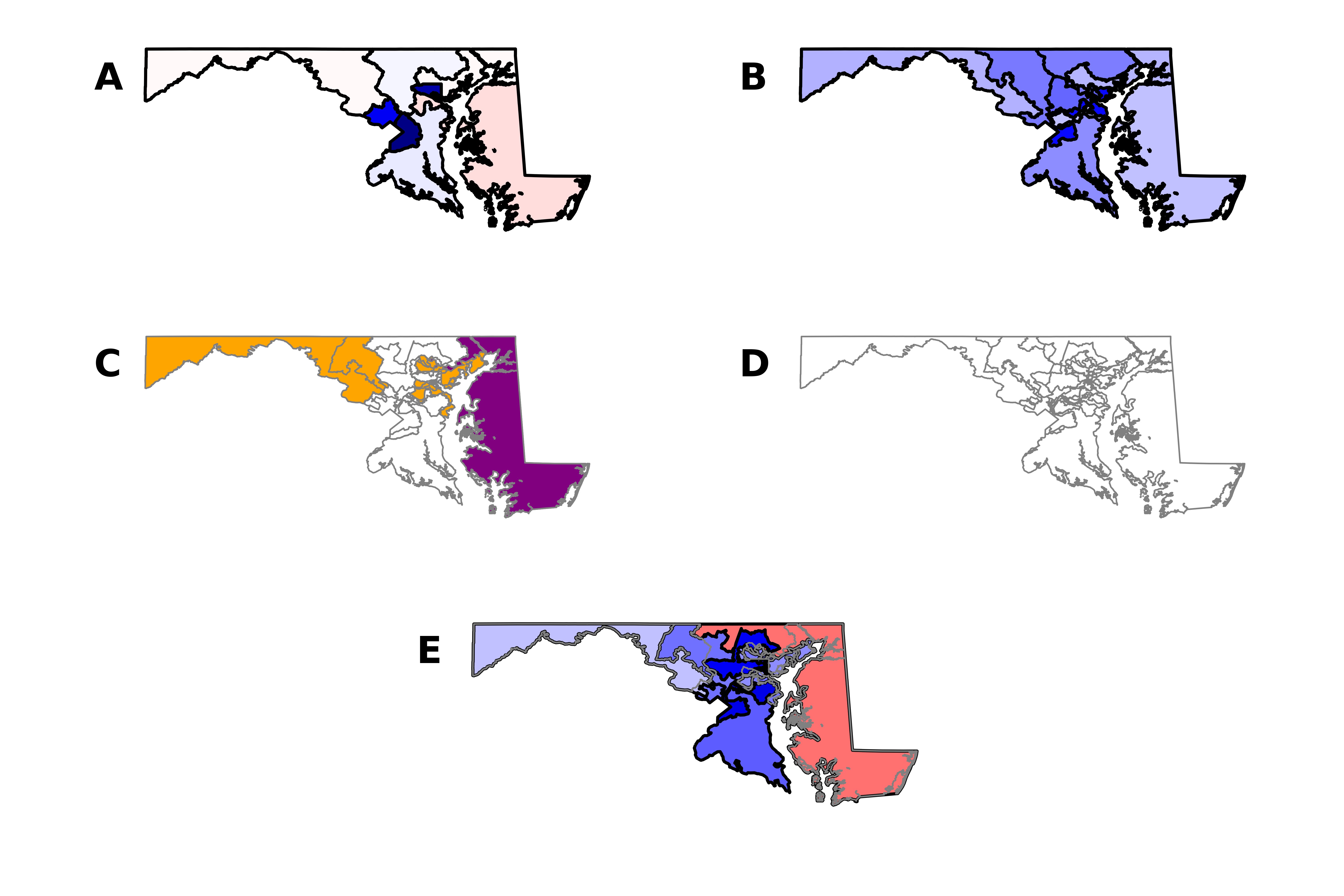}
  \caption{\mycape}
  \label{fig:md}
\end{figure}
}

In Figure~\ref{fig:md}.B we illustrate a hypothetical plan from the
\emph{Atlas} project that maximized the number of Democratic seats. In
this plan, the Democrats are projected to win all seats. So, with this
as a precursor plan, there is no possibility of cracked or packed
Republicans since the Republicans don't win any seats in the precursor
plan. This is illustrated by the lack of either orange or purple
regions in in Figure~\ref{fig:md}.D.

\subsection{North Carolina}
\label{sec:nc}

North Carolina has seen two very recent partisan gerrymandering
cases. \emph{League of Women Voters \emph{v.}\ Rucho} and \emph{Common
  Cause \emph{v.}\ Rucho} both alleged partisan gerrymandering in the
2016 remedial congressional plan. These cases were consolidated in
2017. After a series of maneuvers and developments, the Supreme Court
vacated and remanded a district court decision striking down the map
as an unconstitutional gerrymander. In response to the remand, the court
ordered parties to respond to a number of issues in light of
\emph{Gill \emph{v.}\ Whitford}. The plaintiffs' response focused on
providing additional information to establish standing. They do this
by using a simulated plan, Plan 2--297, generated by plaintiffs' expert
Jowei Chen. In line with the guidance from the Supreme Court, for each
district (except the 3rd), they identify either a plaintiff or a
member of the League of Women Voters of North Carolina who
consistently votes Democratic living in that district.

In this article we identify packed and packed regions of the
\emph{2012} North Carolina congressional plan (i.e., not the plan
under discussion in the \emph{Rucho} cases). Instead of identifying
individual voters we identify packed and cracked VTDs using 1,000
simulated plans from among the 24,000 plans generated
for~\cite{HerschlagNC} and made available by the authors
at~\cite{HerschlagNCdata}. For each voting tabulation district (VTD)
--- comparable for our purposes to a \emph{ward} or \emph{precinct}
--- and each simulated plan used as a comparator, we considered
whether the voters in the VTD had been packed or cracked relative to
the two plans. For this analysis and the one for Wisconsin in
Section~\ref{sec:wi}, we take a more conservative approach than that
taken in the matrix of Figure~\ref{fig:trans-table} and require the
support for a given party to increase or decrease by at least 5 points
before triggering a classification of packed/cracked. Finally, note
that our approach uses the first definition we given in
Section~\ref{sec:pandc} for what it means for a region to be
packed/cracked (i.e., we do not require any sort of mapping of
districts between the comparator plan and the plan of interest).

In Figure~\ref{fig:nc-dem} we illustrate how frequently Democrats,
respectively, in each VTD were characterized as packed or cracked. For
example, the darkest orange (stippled) shading shown indicates that
the Democratic voters in that region were characterized as cracked in
the enacted plan relative to the comparator plans at least 80\% of the
time.

\ifdef{\SUBMIT}{}
{
\begin{figure}
  \centering
  \includegraphics[width=1\linewidth]{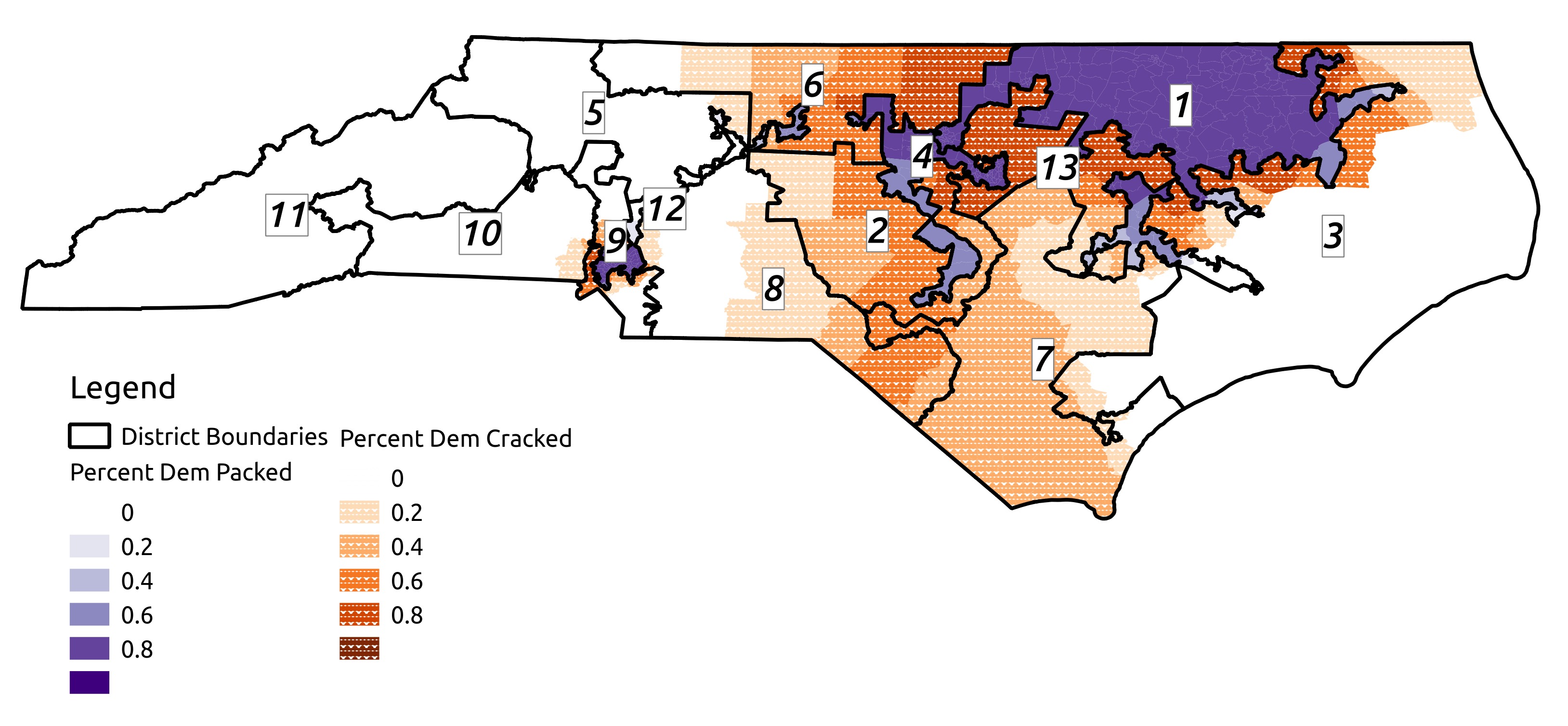}
  \caption{\mycapf}
  \label{fig:nc-dem}
\end{figure}
}

\ifdef{\SUBMIT}{}
{
\begin{figure}
  \centering
  \includegraphics[width=1\linewidth]{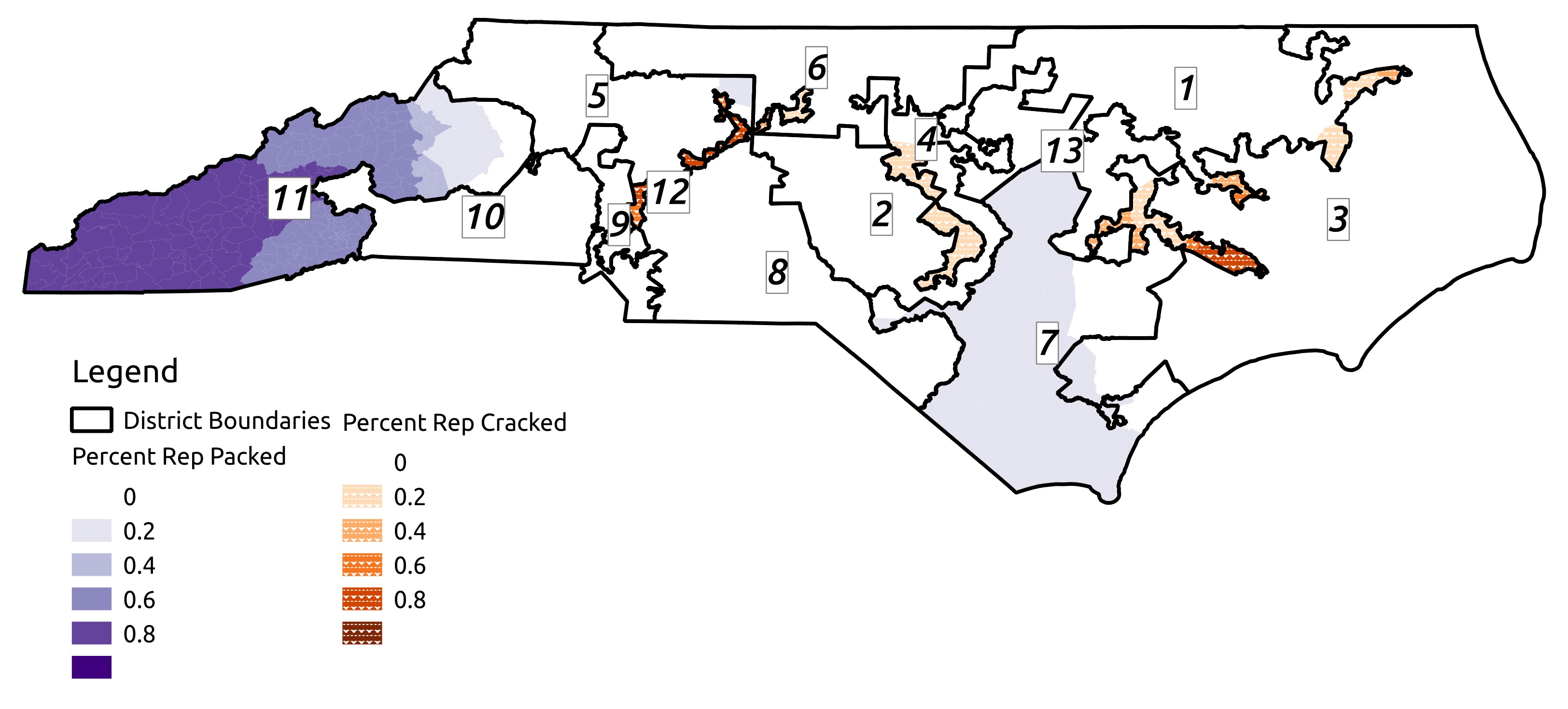}
  \caption{\mycapg}
  \label{fig:nc-rep}
\end{figure}
}

Note that according to the criteria set out in
Section~\ref{sec:pandc}, voters of both parties are injured by
partisan gerrymandering. In light of this, in Figure~\ref{fig:nc-rep}
we illustrate how frequently Republicans were packed or cracked
relative to the comparator plans. According to those criteria, what
matters for standing is not the coverage of packed and cracked VTDs
within a district, but merely the fact that there exist packed and
cracked voters (or VTDs). In fact, for more than 80\% of the computer
simulations, there exists at least one VTD packed with Democrats in
each of the 1st, 4th and 12th congressional districts. The analogous
fact is true for cracked Democratic voters in the 2nd, 3rd, 6th, 8th,
9th and 13th districts. If we include the Republican voters in 11th
who are packed 80\% of the time and the Republican voters in the 1st
(and 12th) who are cracked 80\% of the time, we see that only the 5th,
7th and 10th districts lack packed/cracked voters this
consistently. Note that the plaintiffs' brief does not identify any
packed/cracked voters in the year-2016 3rd district.



From Figure~\ref{fig:nc-rep}, we see that Republican voters in the
western half of the 11th district are packed relative to about half of
the simulated plans. This would be consistent with a partisan
gerrymandering that shores up the 11th district as a relatively safe
Republican seat relative to its natural status (as far as the
computer simulations are concerned) as a swing district. We also see
Republican voters in the 1st and 12th districts who look to have been
cracked. Cracking Republican voters does nothing directly to help
Republicans win more seats. However, there are several reasons while
small populations of Republicans might be moved from Republican
districts into Democratic districts. It might be necessary to equal
out populations. Or, as is more likely the case in these instances,
the narrow strips of Republican-leaning areas are required for
contiguity, connecting disparate areas of Democratic support. The 12th
district is bounded by Democratic areas of Charlotte on the
southwest and the Piedmont Triad of Winston-Salem, Greensboro and
Highpoint on the northeast. Similarly, the arm of the 1st district
filled with cracked Republicans extends down to the city of New Bern.

\subsection{Wisconsin}
\label{sec:wi}

In~\cite{HerschlagWIdata}, the authors generate over 19,000 simulated
districting plans for the 99 districts of the Wisconsin state
assembly. We use a random subset of 1,000 of these plans to use as
comparators for the current, Act 43, Wisconsin state assembly plan
analogously to our analysis of the North Carolina congressional plan
from Section~\ref{sec:nc}. The results are displayed in
Figure~\ref{fig:wi-dem} (for packed/cracked Democrats) and in
Figure~\ref{fig:wi-rep} (for packed/cracked Republicans). As for North
Carolina, to cut down on noise and spurious results, we only counted a
region as having been packed or cracked if the vote values between the
comparator plan and the Act 43 plan differed by at least 5\%.

\ifdef{\SUBMIT}{}
{
\begin{figure}
  \centering
  \includegraphics[width=1\linewidth]{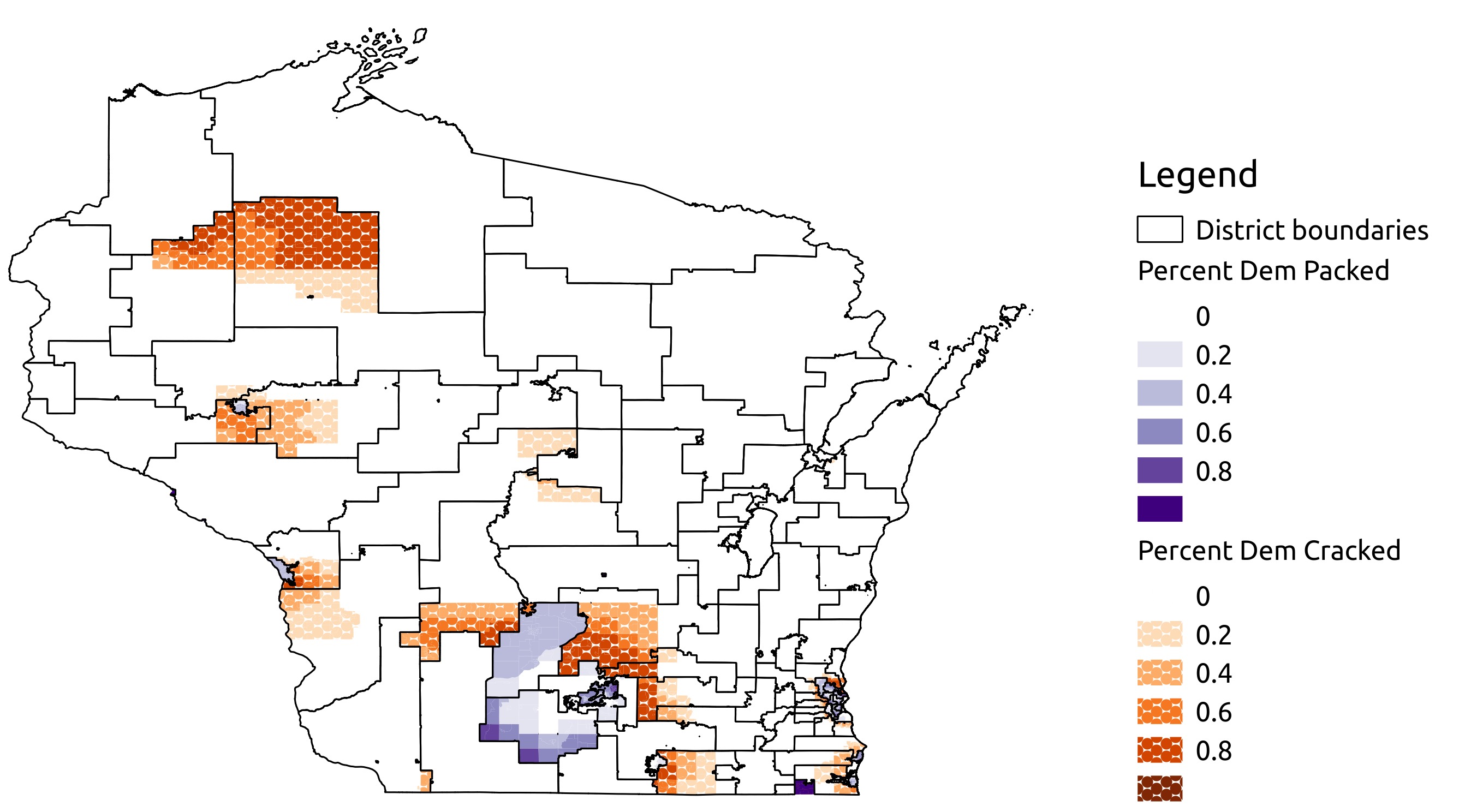}
  \caption{\mycaph}
  \label{fig:wi-dem}
\end{figure}
}

\ifdef{\SUBMIT}{}
{
\begin{figure}
  \centering
  \includegraphics[width=1\linewidth]{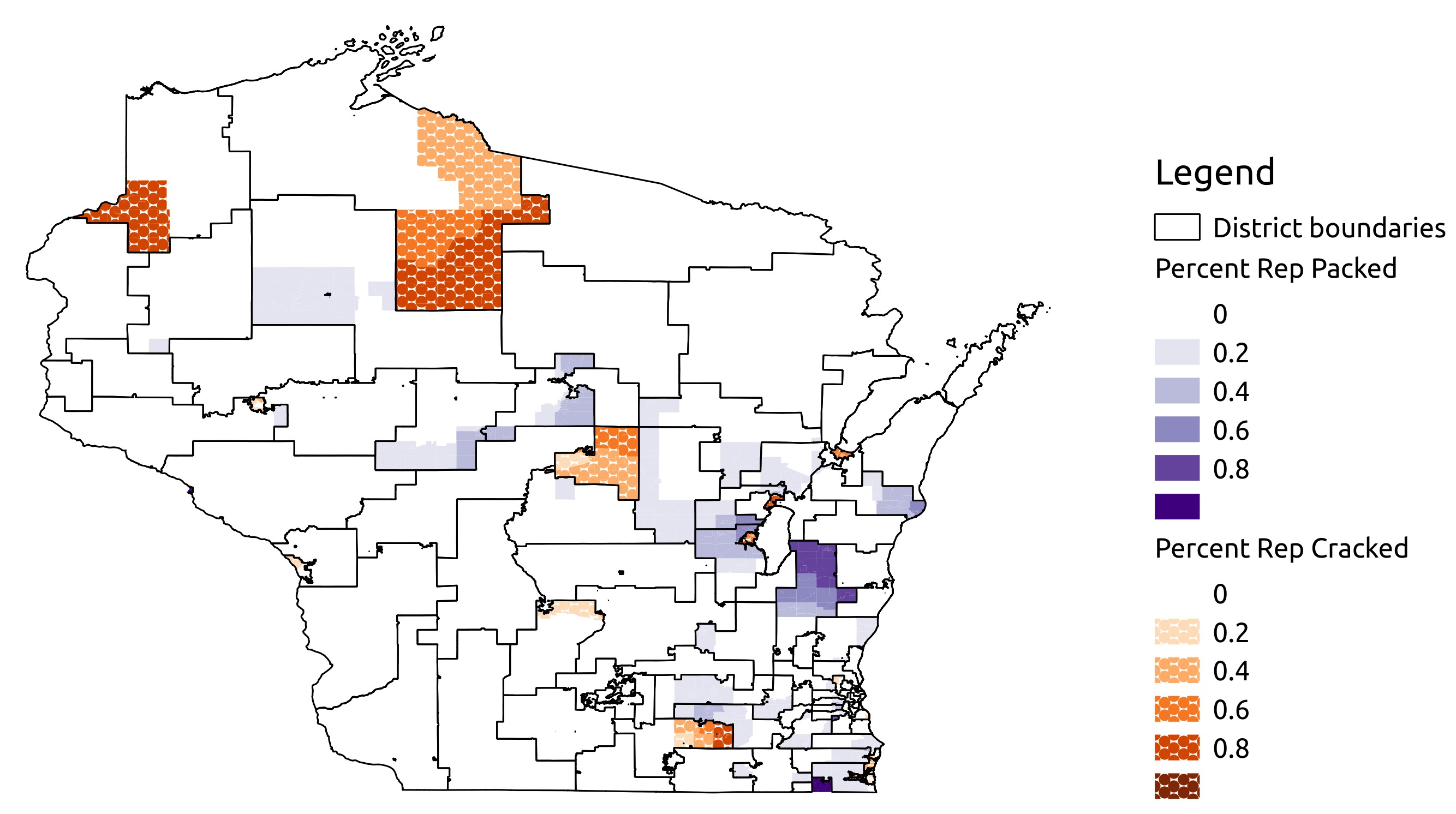}
  \caption{\mycapi}
  \label{fig:wi-rep}
\end{figure}
}

In Table~\ref{tab:dist} we record the frequencies with which each Act
43 district was packed or cracked relative to the 1,000
computer-simulated comparator districts.

\ifdef{\SUBMIT}{}{
\begin{table}[ht]
  \centering
  \caption{Distribution of 99 Wisconsin state legislative districts
    relative to how frequently they contained a packed/cracked
    Democratic/Republican ward. The first column indicates the
    percentage of precursor plans for which a given district at least
    one ward of the indicated type (containing packed Democrats,
    cracked Democrats, etc.). One thousand precursor plans were chosen
    randomly from the data~\cite{HerschlagWIdata} used in~\cite{HerschlagWI}.}
\begin{tabular}{lrrrr}\toprule
  Max frequency &  Dem Packed &  Dem Cracked & Rep Packed & Rep Cracked \\\midrule
  0.0 to 0.19 &   74 &   67 &   68 &   82 \\
  0.2 to 0.39 &    6 &    4 &   19 &    4 \\
  0.4 to 0.59 &    5 &    7 &    7 &    4 \\
  0.6 to 0.79 &    5 &    9 &    3 &    3 \\
  0.8 to 1.00 &    9 &   12 &    2 &    6 \\\bottomrule
\end{tabular}
\label{tab:dist}
\end{table}
}

\section{Discussion}
\label{sec:disc}

There are two basic approaches one can take to identifying a
gerrymander. The first approach is to show that individual districts
have been gerrymandered and, as a result (or by definition), the vote
of certain residents in those districts has been diluted. This
approach seems best suited to when the alleged harm is restricted to a
relatively small portion of the entire district plan, such as the 6th
District and its surroundings in Maryland as alleged in \emph{Benisek
  \emph{v.}\ Lamone}. However, as described by Justice Kagan in her
concurring opinion in \emph{Whitford \emph{v.}\ Gill}, this approach
could also be applied repeatedly in order to address instances in
which the alleged gerrymandering is close to statewide.

However, it is important to note that partisan gerrymanders don't
target individual voters. \emph{Every} district plan will have winners
and losers as measured by who gets to be in the majority and who does
not. The effectiveness of a partisan gerrymander is measured by the
net effect, leading to the second approach of considering a
gerrymander from a statewide perspective from the outset.

The statewide approach is attractive because the districts in a
district plans are typically drawn as parts of a cohesive whole and
because the ultimate goal of a partisan gerrymander is to reduce the
statewide representation of the opposition party. Partisan asymmetry
measures such as the efficiency gap and the declination are well
suited to such a statewide analysis. While such analyses have been
common in the academic literature and popular press, they have had
limited success in the courts. One disadvantage of this approach is
that partisan asymmetry measures seem likely to miss ``localized''
gerrymandering (though perhaps one could apply metrics to only a
portion of the statewide map).

While there are arguments in favor of both approaches, which can be
successful in the courts is ultimately a legal question and one we do
not delve into here. We have had two main goals in this article. The
first has been to make as explicit as possible definitions for, and
some of the pitfalls inherent in, the local approach outlined by both
Chief Justice Roberts and Justice Kagan in their opinions in
\emph{Whitford \emph{v.}\ Gill}. Other definitions for packed/cracked
voters and districts are certainly possible. The second has been to
illustrate how simulations can be used to provide a more robust
picture than can be provided by a single comparator plan, of what has
been packed or cracked. Of course, the cost of using simulations is
that one must take care to show that they have been appropriately
drawn from the universe of possible plans.

\section{Data collection and Methods}

The electoral data and geometry for Maryland used in
Section~\ref{sec:md} are taken from the repository~\cite{atlas}
created as part of the Atlas of Redistricting project. The district
plan shown in Figure~\ref{fig:md}.E is the one used since 2012.

The North Carolina electoral data were obtained from the Harvard
Dataverse~\cite{dataverse-nc}. Partisan lean for each VTD was computed
by averaging the results for the 2012 presidential election and the
North Carolina races for governor, attorney general, treasurer and
secretary of state. The boundaries for congressional districts are
from the Dataverse data and are the boundaries from 2012. These
boundaries were updated in 2016 and hence are \emph{not} the
boundaries being litigated in the \emph{Rucho} cases. The actual
geometries of the VTDs are taken from the Census Bureau~\cite{census-vtd}.
County codes were taken from~\cite{census-county}.

The Wisconsin ward and state Assembly district geometries were
obtained from~\cite{wisconsin-shape}. Ward boundaries are for 2017,
district boundaries are the Act 43 districts used in the 2012--2016
elections. Electoral data (including fixed ward-district associations)
were taken from~\cite{HerschlagWIdata}. Partisan lean for each ward
was computed by averaging the results for six elections: the 2016
presidential and US senatorial elections, the 2014 US senatorial
election and the 2014 gubernatorial, secretary-of-state, and treasurer
elections.

The data were analyzed using python code~\cite{dilution-code} written
by the author in a Jupyter notebook environment~\cite{jupyter}. Python
packages utilized were Pandas~\cite{pandas},
GeoPandas~\cite{geopandas}, Matplotlib~\cite{matplotlib},
NumPy~\cite{numpy} and Shapely~\cite{shapely}. The depictions of
packed and cracked votes for North Carolina and Wisconsin illustrated
in Figures~\ref{fig:nc-dem} to~\ref{fig:wi-rep} were created using
QGis~\cite{qgis}.

\section{Acknowledgments}

The author is grateful to the authors of~\cite{HerschlagNC}
and~\cite{HerschlagWI} for making their collection of
computer-simulated district plans for North Carolina and Wisconsin
publicly available in a format usable by other researchers; the author
would like to thank Greg Herschlag in particular for his help in
accessing these plans.

\ifdef{\SUBMIT}{
\emph{Remaining acknowledgments are redacted.}
}
{
This work was partially supported by a grant from the Simons
Foundation (\#429570). 
}

\bibliography{gerrymandering}

\bibliographystyle{alpha}

\ifdef{\SUBMIT}{
  
\clearpage

\begin{figure}
  \centering
  \includegraphics[width=0.25\linewidth]{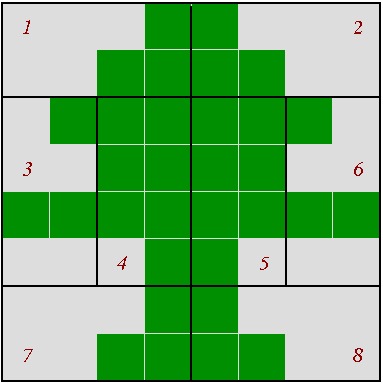}
  \caption{\mycapa}
  \label{fig:ex-single}
\end{figure}

\clearpage

\begin{figure}
  \centering
  \includegraphics[width=0.3\linewidth]{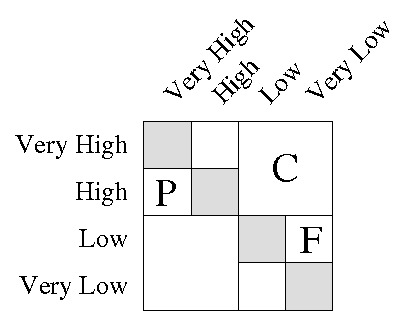}
  \caption{\mycapb}
  \label{fig:trans-table}
\end{figure}

\clearpage

\begin{figure}
  \centering
  \includegraphics[width=0.7\linewidth]{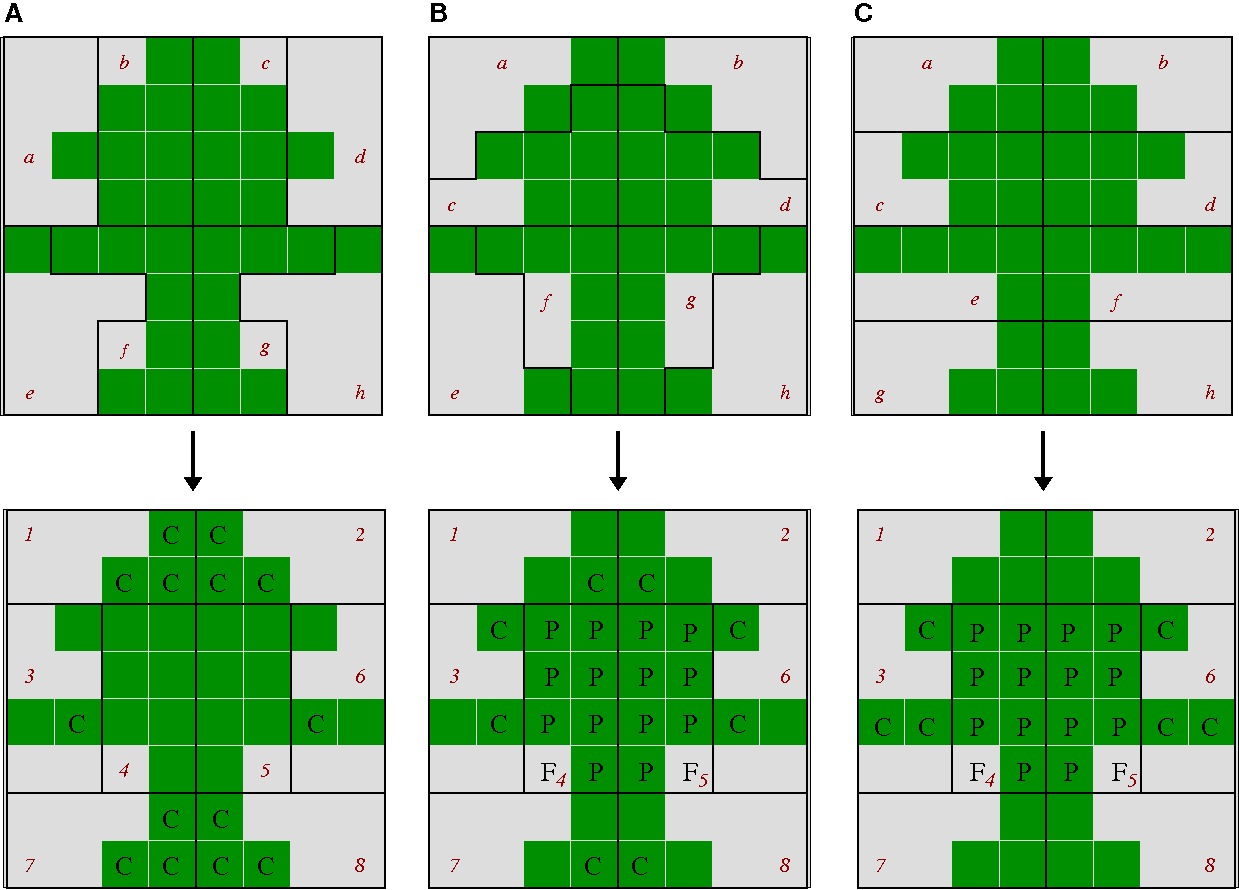}
  \caption{\mycapc}
  \label{fig:ex-pandc}
\end{figure}

\clearpage

\begin{figure}
  \centering
  \includegraphics[width=1\linewidth]{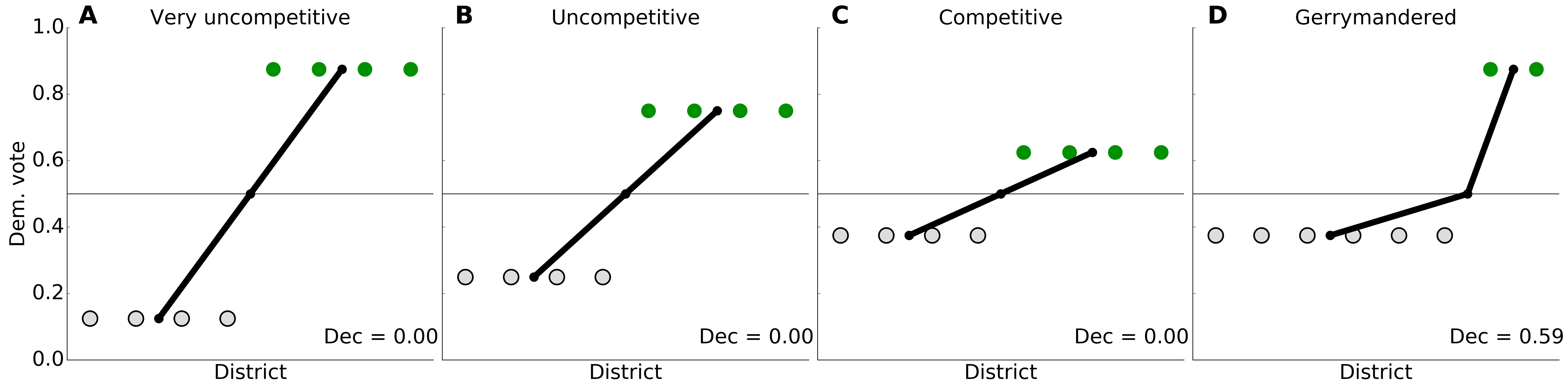}
  \caption{\mycapd}
  \label{fig:prec-vd}
\end{figure}

\clearpage

\begin{figure}
  \centering
  \includegraphics[width=1\linewidth]{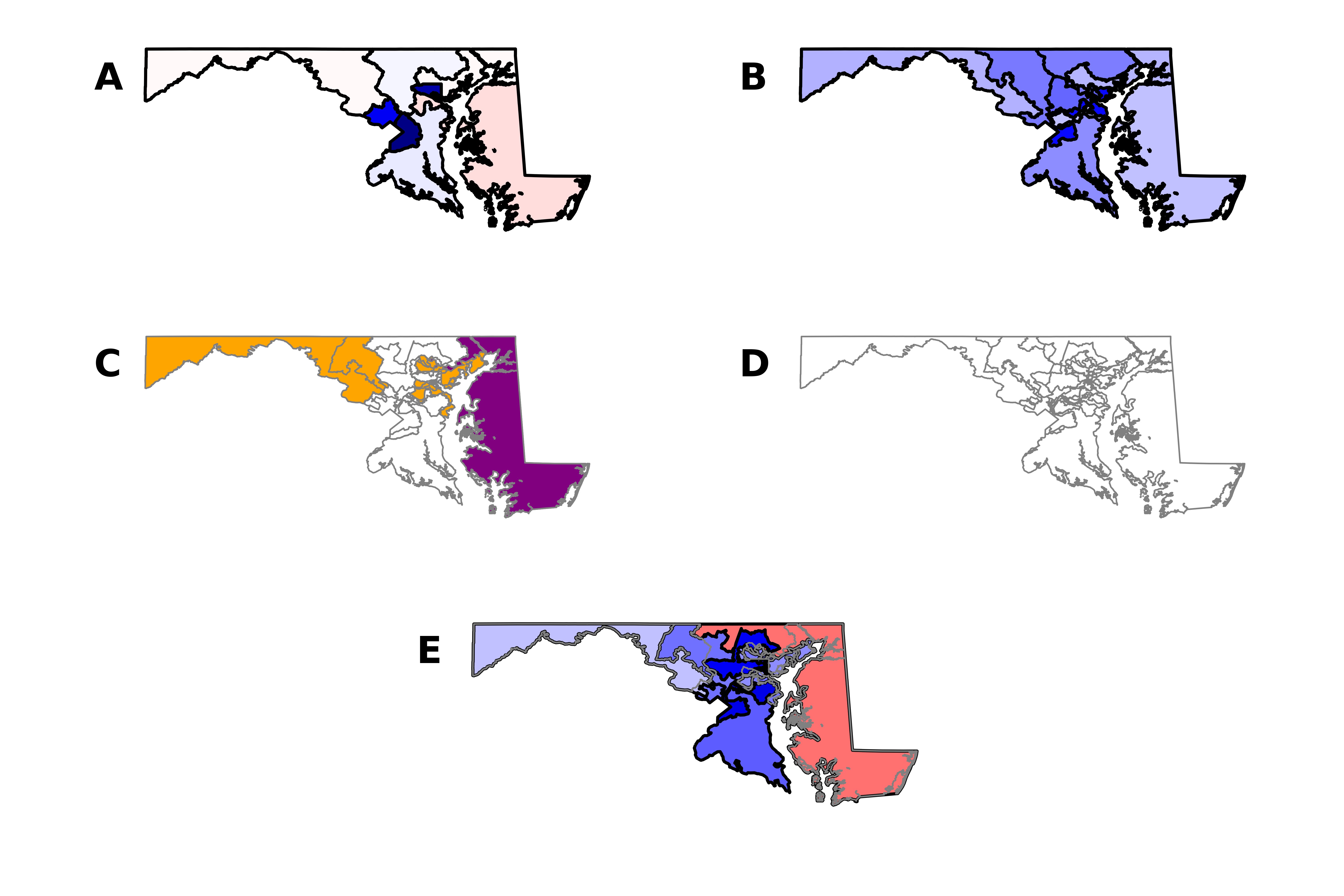}
  \caption{\mycape}
  \label{fig:md}
\end{figure}

\clearpage

\begin{figure}
  \centering
  \includegraphics[width=1\linewidth]{NC-Dem-pandc-aug01-crop}
  \caption{\mycapf}
  \label{fig:nc-dem}
\end{figure}

\clearpage

\begin{figure}
  \centering
  \includegraphics[width=1\linewidth]{NC-Rep-pandc-aug01-crop}
  \caption{\mycapg}
  \label{fig:nc-rep}
\end{figure}

\clearpage

\begin{figure}
  \centering
  \includegraphics[width=1\linewidth]{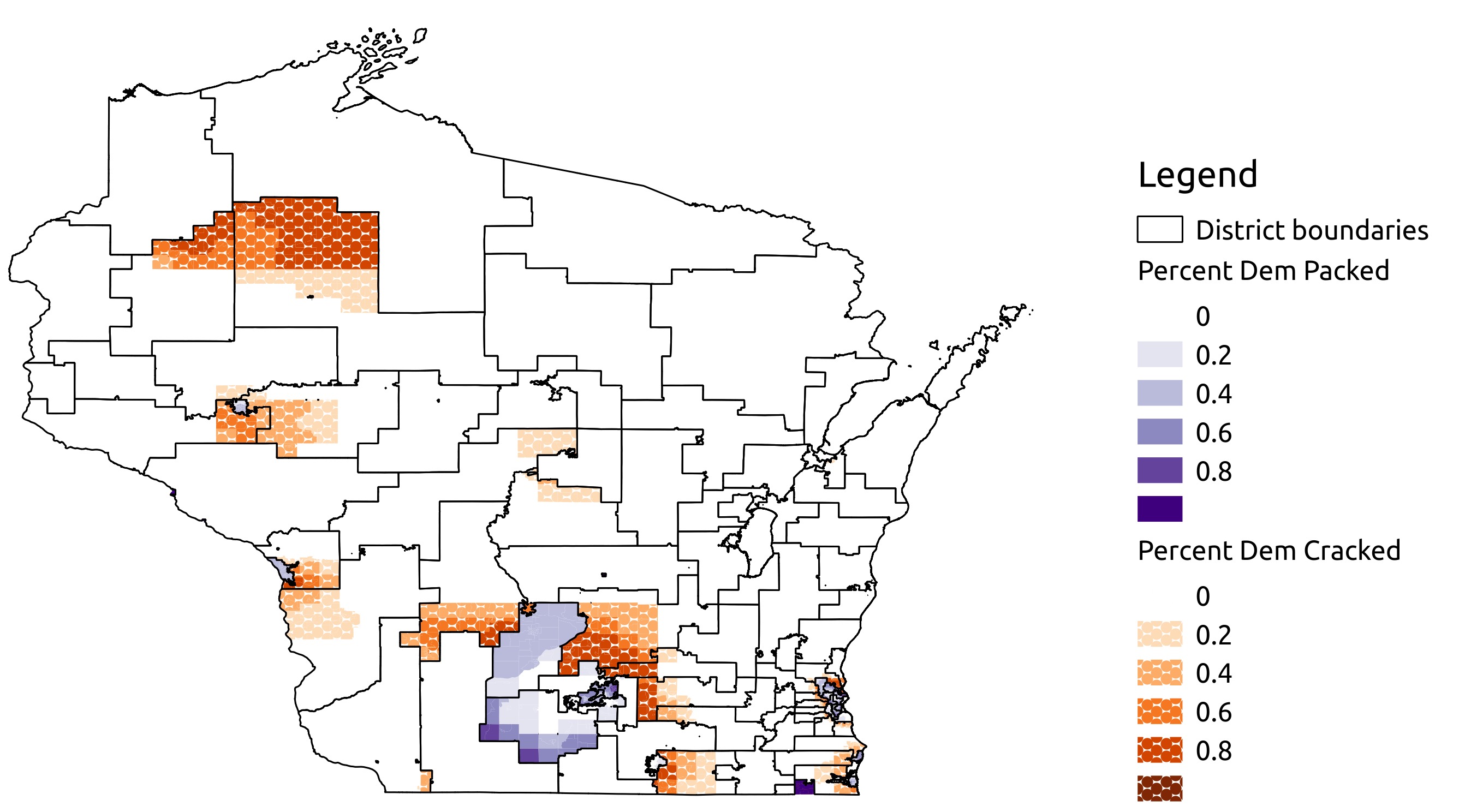}
  \caption{\mycaph}
  \label{fig:wi-dem}
\end{figure}

\clearpage

\begin{figure}
  \centering
  \includegraphics[width=1\linewidth]{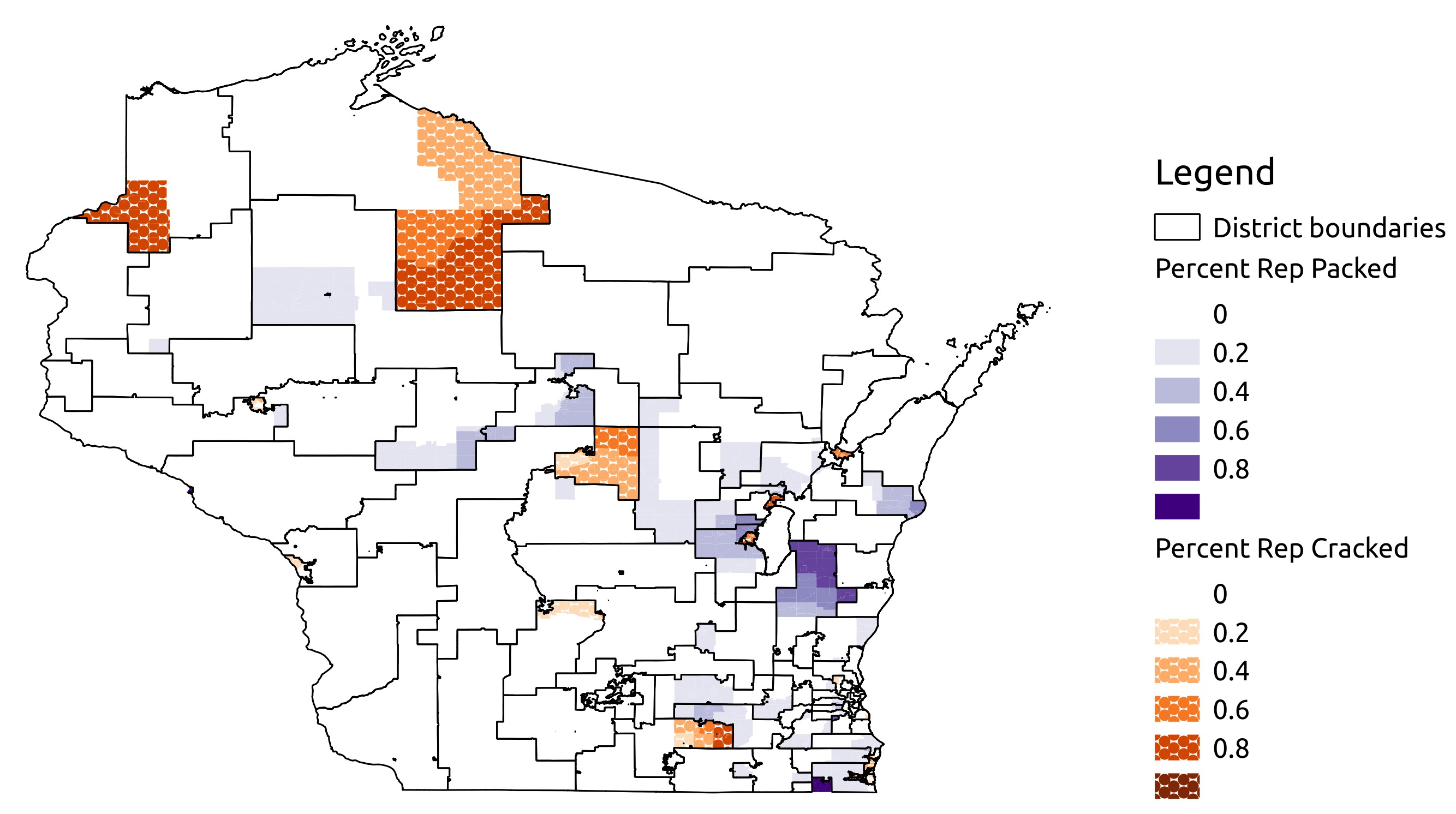}
  \caption{\mycapi}
  \label{fig:wi-rep}
\end{figure}

\clearpage

\begin{table}[ht]
  \centering
  \caption{Distribution of 99 Wisconsin state legislative districts
    relative to how frequently they contained a packed/cracked
    Democratic/Republican ward. The first column indicates the
    percentage of precursor plans for which a given district at least
    one ward of the indicated type (containing packed Democrats,
    cracked Democrats, etc.). One thousand precursor plans were chosen
    randomly from the data~\cite{HerschlagWIdata} used in~\cite{HerschlagWI}.}
\begin{tabular}{lrrrr}\toprule
  Max frequency &  Dem Packed &  Dem Cracked & Rep Packed & Rep Cracked \\\midrule
  0.0 to 0.19 &   74 &   67 &   68 &   82 \\
  0.2 to 0.39 &    6 &    4 &   19 &    4 \\
  0.4 to 0.59 &    5 &    7 &    7 &    4 \\
  0.6 to 0.79 &    5 &    9 &    3 &    3 \\
  0.8 to 1.00 &    9 &   12 &    2 &    6 \\\bottomrule
\end{tabular}
\label{tab:dist}
\end{table}

\clearpage

\pagestyle{empty}

\clearpage

{\bf Figure 1: }\mycapa\\

{\bf Figure 2: }\mycapb\\

{\bf Figure 3: }\mycapc\\

{\bf Figure 4: }\mycapd\\

{\bf Figure 5: }\mycape\\

{\bf Figure 6: }\mycapf\\

{\bf Figure 7: }\mycapg\\

{\bf Figure 8: }\mycaph\\

{\bf Figure 9: }\mycapi\\

}

\end{document}